\documentclass[english,twocolumn,superscriptaddrconfigurationess,aps,
  longbibliography]{revtex4-1}

\usepackage{soul}
\usepackage[dvipsnames]{xcolor}
\usepackage[utf8]{inputenc}
\usepackage[T1]{fontenc}
\usepackage{xcolor}
\usepackage{graphicx}
\usepackage{amsmath}
\usepackage{amsfonts}
\usepackage{color}
\usepackage{babel}
\usepackage{url}
\usepackage{hyperref}
\usepackage{tikz}
\usepackage{CJKutf8}
\usepackage[title]{appendix}
\usepackage{siunitx}
\usepackage{braket}
\usepackage{tabularx}

\definecolor{lime}{HTML}{A6CE39}
\DeclareRobustCommand{\orcidicon}{
  \begin{tikzpicture}
  \draw[lime, fill=lime] (0,0) 
  circle [radius=0.16] 
  node[white] {{\fontfamily{qag}\selectfont \tiny ID}};
  \draw[white, fill=white] (-0.0625,0.095) 
  circle [radius=0.007];
  \end{tikzpicture}
  \hspace{-2mm}
}

\foreach \x in {A, ..., Z}{\expandafter\xdef\csname
  orcid\x\endcsname{\noexpand\href{https://orcid.org/\csname
      orcidauthor\x\endcsname} {\noexpand\orcidicon}} }

\begin{document}

\title{Sextupole reduction via chaos suppression at the National Synchrotron Light Source II}

\author{Yongjun Li\orcidA{}}\thanks{email: yli@bnl.gov}
\affiliation{Brookhaven National Laboratory, Upton, New York 11973, USA}
\author{Minghao Song}\affiliation{Brookhaven National Laboratory, Upton, New York 11973, USA}
\author{Yoshiteru Hidaka}\affiliation{Brookhaven National Laboratory, Upton, New York 11973, USA}
\author{Victor Smaluk}\affiliation{Brookhaven National Laboratory, Upton, New York 11973, USA}
\author{Timur Shaftan}\affiliation{Brookhaven National Laboratory, Upton, New York 11973, USA}

\begin{abstract}
  We revisit the nonlinear lattice design approach for the National Synchrotron Light Source II (NSLS-II) storage ring. By suppressing chaos, we identify alternative sextupole configurations to the original design, which relied on the conventional strategy of simultaneously minimizing Resonance Driving Terms (RDT) and Amplitude-Dependent Detuning (ADD). These alternatives achieve comparable performance while requiring fewer sextupoles. A detailed comparison of two representative solutions is presented and supported by experimental validation. Our results show that the dynamic aperture correlates more strongly with global chaos than with individual RDTs, and that the importance of minimizing ADD may have been overstated in earlier design strategies.
\end{abstract}

\maketitle

\section{Introduction}
  When designing ring-based accelerators, their linear lattices are optimized to achieve key performance goals for users -- such as luminosity for colliders and brightness for light sources. Subsequently, nonlinear magnets are incorporated to address operational requirements, including chromaticity compensation, injection efficiency, beam lifetime, and other operational requirements. The presence of nonlinear magnets typically renders accelerators non-integrable Hamiltonian systems. As a result, exact solutions for particle motion are generally unattainable due to the lack of sufficient conserved quantities. However, long-term stable motion can be maintained within certain regions of the six-dimensional phase space based on the Kolmogorov-Arnold-Moser theory~\cite{de2001tutorial}. Within these regions, quasi-periodic motions exist on approximate invariant tori, which are partially distorted to resist chaotic diffusion.

  In accelerator physics, the stable region around the origin is referred to as the Dynamic Aperture (DA)~\cite{chao2023handbook}. Its projection onto the transverse plane at the injection point determines whether an injected beam can be successfully captured. The maximum allowable relative momentum deviation ($\Delta P / P_0$) along the longitudinal direction, known as the Local Momentum Aperture (LMA), governs the probability of particle survival following intra-beam scattering. Thus, achieving sufficient DA and LMA is a fundamental requirement in lattice design.

  At National Synchrotron Light Source II (NSLS-II), our previous design philosophy was that, by controlling specific nonlinear performance parameters -- including Resonance Driving Terms (RDTs), Amplitude-Dependent Detuning (ADD), and chromaticity -- we could effectively push the onset of chaotic motion to larger amplitudes. Since nonlinear dynamics cannot be fully described analytically, numerical simulations remain indispensable for confirming long-term stability. However, long-term tracking is both computationally intensive and provides only limited physics insight into the underlying dynamics. The development of more efficient physics-guided lattice design approaches remains an active area of research. Various Chaos Indicators (CI) have been utilized as the optimization objectives in recent years. In this paper, we present a comprehensive comparison of two representative solutions obtained using different approaches: minimizing RDTs and suppressing chaos.

\section{Resonance driving terms and chaos indicators}

  In a storage ring, a single particle's motion is governed by the Hamiltonian $H$, which is defined by the accelerator lattice. $H$, when expanded, has terms like:
  $$H = H_2 + H_{n\ge3},$$
  where $H_2$ represents all quadratic monomial terms to represent the linearized motion. The higher order terms $H_{n\ge3}$ are mainly generated by nonlinear magnetic fields such as sextupoles, octupoles, and other nonlinear magnets. When nonlinear effect is sufficiently weak, through a normal form analysis ~\cite{dragt1982lectures,chao2020lectures}, $H_{n\ge3}$ terms are formulated as perturbations to $H_2$~\cite{wang2012explicit}. When the tune satisfies a resonance condition, the corresponding RDT quantifies how strongly the resonance is being driven. The geometric terms are often related to on-momentum DA, and the chromatic terms affect the off-momentum motion and ultimately determine the LMA. Even when the linear tune is far from a specific resonance, ADD terms can push the tune to match the resonant condition at certain amplitudes, driving the particle motion to be chaotic or even unbounded. The same mechanism also applies to off-energy particles with energy-dependent detuning.

  Since each specific RDT can potentially excite a corresponding nonlinear resonance, it is desirable to minimize multiple RDTs simultaneously -- particularly those of lower orders. However, in practice, only a limited number of them can be effectively controlled due to the finite number of available tuning knobs. Even so, higher-order RDTs that remain uncontrolled may still have detrimental effects on the dynamic aperture. In other words, only minimizing low-order RDTs is a necessary but not sufficient condition for achieving the desired DA~\cite{yang2011multiobjective}. As a result, comprehensive tracking simulations remain indispensable for evaluating long-term stability.

  Besides the perturbation approaches, various CI~\cite{bazzani2023, montanari2025chaos} -- both experimental and numerical -- have been adopted to detect and quantify chaotic behavior in accelerators. These CIs can distinguish regular and chaotic particle motion. Commonly used CIs include, though not limited to, Lyapunov exponents~\cite{benettin1980lyapunov}, Frequency Map Analysis (FMA)~\cite{papaphilippou2014}, forward-reversal integration~\cite{panichi2017, li2021fast}, entropies~\cite{li2024approximate, li2025online}, and others. Unlike approaches that target specific RDTs in the Hamiltonian, the CI methods aim to enhance the regularity of particle motion by suppressing global chaos, thereby enlarging the DA. Quantitative assessments of chaos are typically derived from particle trajectories in phase space, particularly through Poincar\'e maps or spectral analysis. At facilities such as NSLS-II and APS-U~\cite{sun2017comparison}, optimization by pushing the onset of chaotic motion to larger amplitudes has been proven effective in maximizing their DAs.

\section{Evolution of nonlinear lattice optimization at NSLS-II}

\begin{figure}
  \centering
  \includegraphics[width=0.9\columnwidth]{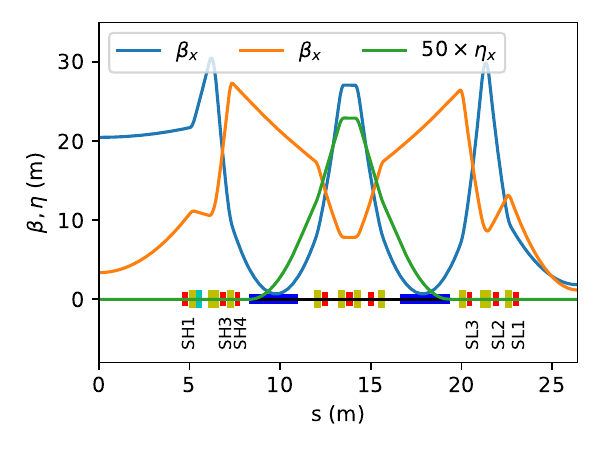}
  \caption{Magnet layout and linear optics of one cell in the NSLS-II storage ring. Sextupoles are shown as red blocks. Six families of harmonic sextupoles, annotated in the non-dispersive sections, are used to compensate nonlinearities introduced by the three families of chromatic sextupoles.} 
  \label{fig:nsls2cell}
  \end{figure}

  The NSLS-II, a third-generation light source, features an emittance of \SI{2}{nm} and a double-bending achromat lattice. Its storage ring consists of 15 identical supercells, each composed of two mirror-symmetric single cells, as illustrated in Fig.~\ref{fig:nsls2cell}. For nonlinear tuning, the machine employs six families of harmonic sextupoles and three families of chromatic sextupoles. The lattice design process, initiated in 2006 and finalized around 2012, focused on minimizing first- and second-order RDTs, i.e., $H_{a+b+c+d+e=3,4}$, and on evaluating the DA and LMA through tracking simulations. Given the large number of RDTs, configurations with up to nine families of harmonic sextupoles were explored, but ultimately six were adopted. The final configuration was selected after rigorous robustness checks that accounted for engineering imperfections and insertion device integration.
  
  In recent years, different approaches to using CIs have been explored.  Rather than focusing on each individual resonance, we aim at controlling the overall nonlinear lattice performance characterized with various CIs~\cite{li2021fast, li2021design, li2022data, li2024approximate, li2025online, li2025construction}. Among these, a representative solution obtained through suppressing fluctuations of Approximate Invariants (AI)~\cite{li2025construction} has been selected for comparison with our original design.

\section{Performance comparison of nonlinear dynamics\label{comparison}}

  In this section, we present a detailed performance comparison between these two representative solutions. The original design is referred to as the Resonance Driving Term Minimization (RDTM) configuration, while the recently re-optimized solution is designated the Chaos Suppression (CS) configuration. Throughout this paper, unless otherwise specified, the properties of the RDTM configuration are shown in the left columns of figures and tables, and those of the CS configuration in the right columns.
  
\subsection{Sextupole configurations}
  
  Sextupole strengths, defined as:
$$K_{2}=\frac{1}{(B\rho)_0 }\frac{\partial ^{2}B_{y}}{\partial x^{2}},$$
with a given beam rigidity $(B\rho)_0$, are provided for the two configurations in Table \ref{tab:k2}. Notably, the \textit{SH4} family is entirely disabled in the CS configuration. This highlights a key difference: while the RDTM scheme necessitates more sextupole knobs to control multiple objectives, the new approach aims at suppressing global chaos parameterized as a single objective, achieving the goal with a reduced number of tuning knobs.

  \begin{table}[ht]
  \centering
  \begin{tabular*}{0.8\columnwidth}{@{\extracolsep{\fill}} c|r|r}
  \hline\hline
         & RDTM & CS \\
  \hline
        Sext. name & $K_2\;(\mathrm{m}^{-3})$ & $K_2\;(\mathrm{m}^{-3})$ \\
  \hline
        SH1 & 19.83 & 26.80 \\
        SH3 & -5.86 & -21.89 \\
  \textit{SH4} & \textit{-15.82} & \textit{0} \\
        SL1 & -13.27 & -2.10 \\
        SL2 & 35.68 & 30.80 \\
        SL3 & -29.46 & -23.00 \\
  \hline
  \end{tabular*}
  \caption{Strengths of the harmonic sextupoles in Fig.~\ref{fig:nsls2cell} for two configurations. The \textit{SH4} family is disabled in the CS configuration.}
  \label{tab:k2}
  \end{table}
      
\subsection{Poincar\'e map}

\begin{figure}
  \centering
  \includegraphics[width=0.45\columnwidth]{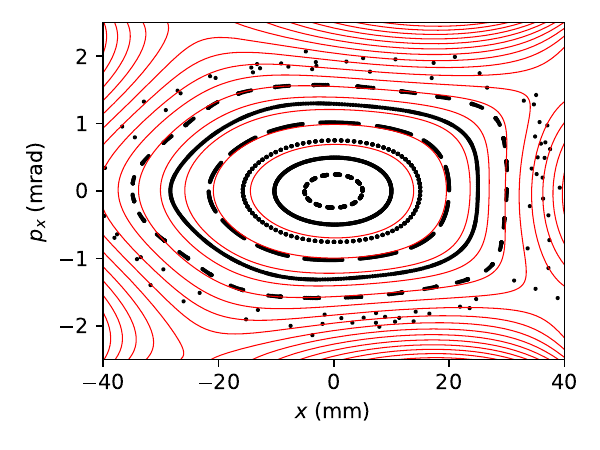}
  \includegraphics[width=0.45\columnwidth]{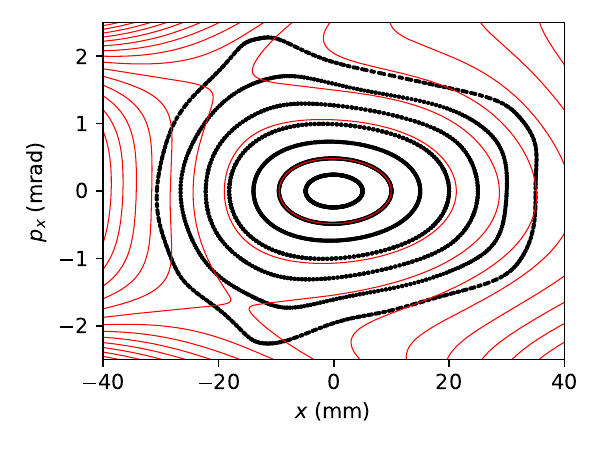}
  \includegraphics[width=0.45\columnwidth]{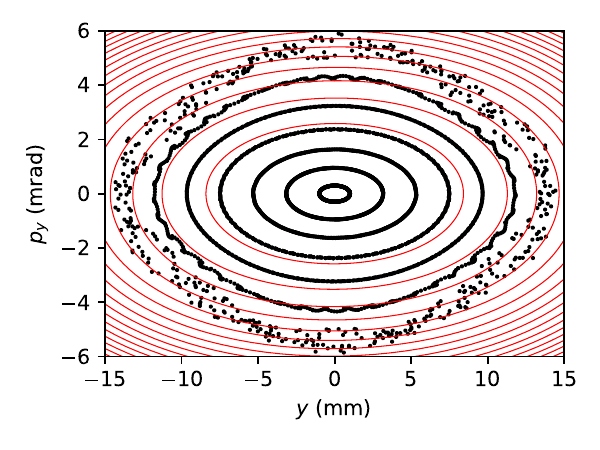}
  \includegraphics[width=0.45\columnwidth]{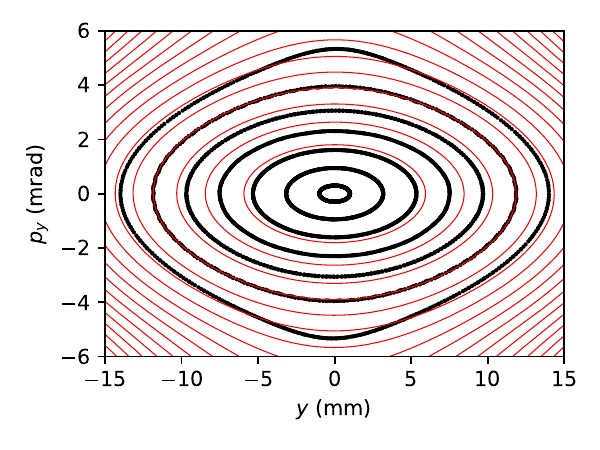}
  \caption{The Poincar\'e maps observed at the injection point for the two configurations in the horizontal (top) and vertical (bottom) planes. Background red contour lines represent their $5^{th}$-order approximate invariant tori; black dots are simulated turn-by-turn trajectories.} 
  \label{fig:poincare}
  \end{figure}

  In quasi-periodical nonlinear dynamical systems, Poincar\'{e} maps -- constructed from the intersections of multi-turn phase-space trajectories with a designated transverse cross-section -- offer not only a visual representation but also a quantitative measure of chaos. In accelerator physics, such maps can be generated through either numerical simulations or experimental measurements~\cite{li2024dedicated}. Fig.~\ref{fig:poincare} presents simulated transverse Poincar\'{e} maps for the two configurations under study. In the RDTM configuration, the horizontal Poincar\'{e} map exhibits a pentagonal shape with a tip oriented toward the inboard side, resulting in an asymmetric projected $x-y$ aperture, as shown in Fig.~\ref{fig:dyap}. In contrast, the CS configuration demonstrates greater momentum acceptance, corresponding to a larger occupied phase-space volume. In the vertical plane, the RDTM configuration shows clear signs of chaotic motion at large amplitudes, a feature that is notably absent in the CS configuration.

\subsection{Dynamic aperture}
\begin{figure}
  \centering
  \includegraphics[trim=0 0 50 0,clip,width=0.9\columnwidth]{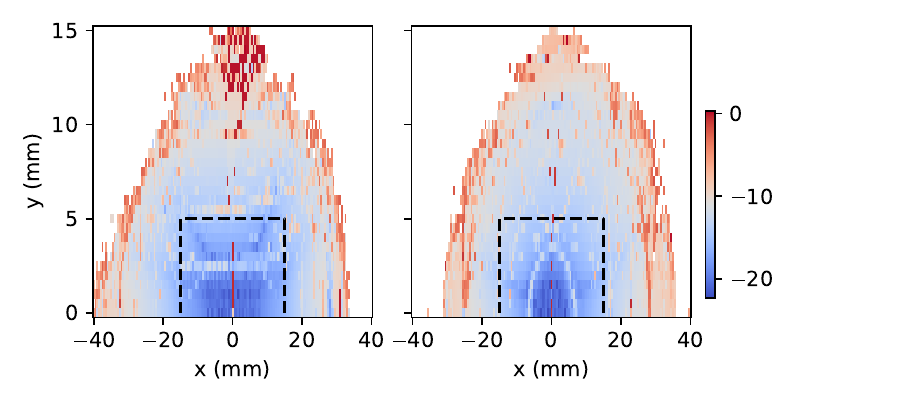}
  \caption{The projections of DA in the transverse $x-y$ plane (on-momentum dynamic aperture) at the injection point. The black dashed boxes specify the desired dimension ($\pm15\times\pm5\si{mm}$). The colors here and also in following Fig.~\ref{fig:xd} and \ref{fig:fma} are the chaos measured with tune diffusion $\log_{10}\sqrt{\Delta\nu_x^2+\Delta\nu_y^2}$~\cite{laskar1999introduction} .}
  \label{fig:dyap}
  \end{figure}

  In accelerator physics, the DA is commonly defined by its 2D projection in the transverse $x-y$ plane at the injection point. A sufficient aperture is required to accommodate the incident beam, accounting for effects such as finite bunch size, energy spread, magnet misalignment, and other lattice imperfections. As demonstrated in Fig.~\ref{fig:dyap}, the DAs for both configurations are sufficient, with the desired dimensions denoted by the black dashed boxes. In this figure, the colors represent tune diffusion~\cite{laskar1999introduction}, which will be explained in detail later in the text.
    
\begin{figure}
  \centering
  \includegraphics[trim=0 0 50 0, clip, width=0.9\columnwidth]{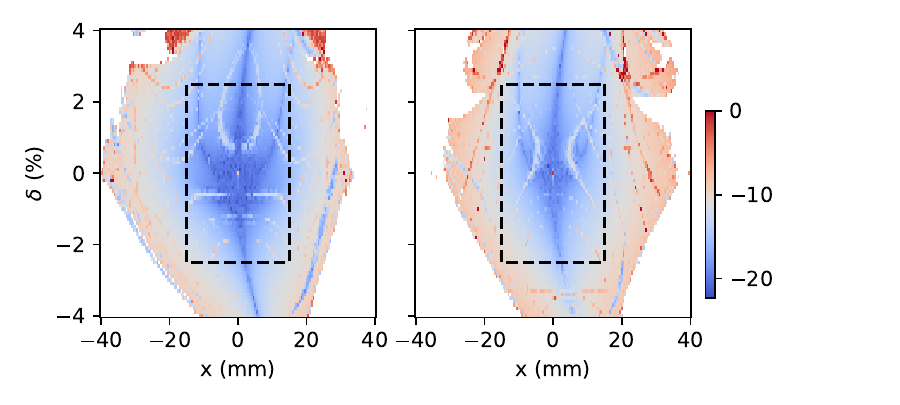}
  \caption{The projections of DA in the $x-\delta$ plane. The black dashed boxes define the desired dimension ($\pm15\si{mm}\times\pm2.5\%$).} 
  \label{fig:xd}
  \end{figure}

  The 2D projection in the horizontal-momentum ($x-\delta$) plane is a critical diagnostic used to visualize the maximum allowed aperture for both off-axis and off-energy particles at the injection point. This comparison is shown in Fig.~\ref{fig:xd}.
  
\subsection{Frequency map analysis and amplitude-dependent detuning}

  FMA projects the amplitude-dependent tune footprint, which is colored according to its diffusion~\cite{laskar1999introduction}. Tune diffusion is defined as the fluctuation of the tune observed over different segments of the same trajectory. It represents the sensitivity of the betatron tune to initial conditions and precisely identifies tune smear in the vicinity of resonances. Tune diffusion is also considered a chaos indicator extracted through precise spectral analyses~\cite{papaphilippou2014}.

\begin{figure}
  \centering
  \includegraphics[trim=0 0 50 0, clip,width=0.9\columnwidth]{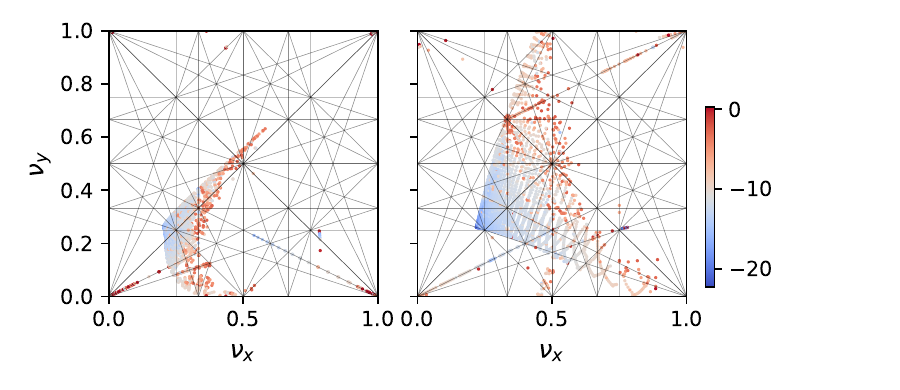}
  \caption{The FMA in the tune space, also referred to as tune footprint, for two configurations. A relatively larger ADD in the CS configuration doesn't degrade its DA.} 
  \label{fig:fma}
  \end{figure}

\begin{figure}
  \centering
  \includegraphics[width=0.9\columnwidth]{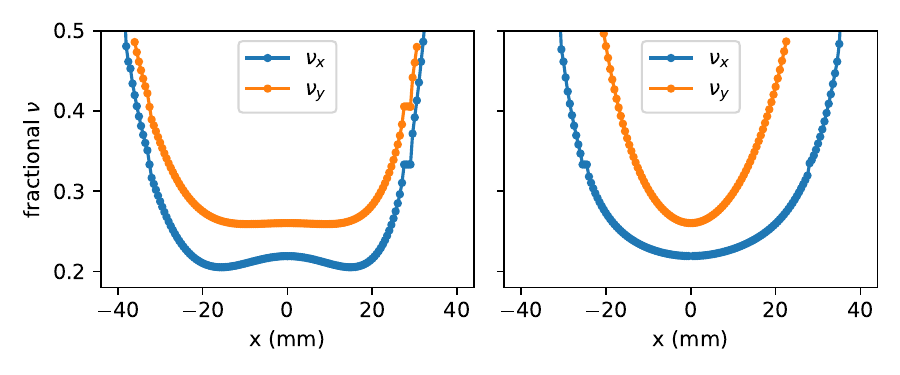}
  \caption{The ADD in the horizontal plane ($y=0$) for two configurations.} 
  \label{fig:tswa}
  \end{figure}

  The FMAs for the two configurations, as illustrated in Fig.~\ref{fig:fma}, have distinct patterns. In the RDTM configuration, the linear ADDs are well minimized, and the tune footprint is confined within a small range. In contrast, in the CS configuration, well-controlled fluctuations in the approximate invariants allow the tune to safely cross low-order resonances, even $\nu_x=\frac{1}{3}$. Detailed ADD information in the horizontal plane is also illustrated in Fig.~\ref{fig:tswa}.

  From these comparisons of the tune footprint in FMA and ADD, we conclude that chaos suppression pushes the transition from regular to chaotic motion to large oscillation amplitude, causing the lattice to behave more like a near-integrable system. This enhances its resistance to chaos and allows for safer resonance crossing. Similar resonance-crossing behavior has been observed in Integrable Optics Test Accelerator (IOTA) machines~\cite{nagaitsev2013beam,ruisard2019single,hwang2020rapidly}. Traditionally, enlarging the dynamic aperture has relied on simultaneously avoiding resonance crossing and narrowing the resonance stop-band width. This conventional approach therefore places excessive emphasis on minimizing ADD. In contrast, the CS configuration suggests an alternative approach: optimizing the lattice to be as close as possible to an integrable system, potentially granting it similar resonance-crossing robustness as observed in integrable systems. It is also worth noting that although the tune footprint appears similar to the solution reported in Ref.~\cite{li2021design}, the corresponding sextupole strengths differ significantly, further indicating that the solutions are not unique.

\subsection{Longitudinal variation of RDTs}

\begin{figure}
  \centering
  \includegraphics[width=0.9\columnwidth]{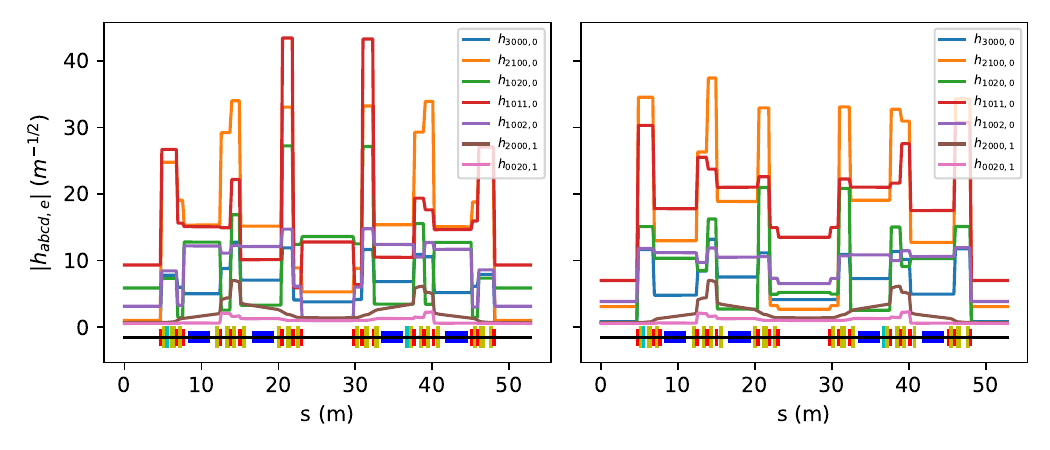}
  \includegraphics[width=0.9\columnwidth]{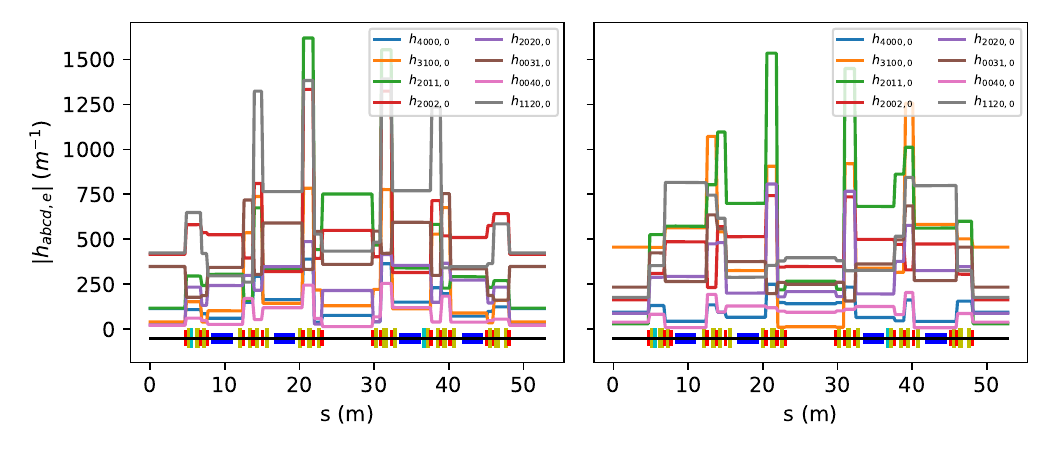}
  \caption{The amplitude variation of the third-order (top) and fourth-order (bottom) RDTs along the longitudinal direction for two configurations.} 
  \label{fig:h3h4}
  \end{figure}

  Previous studies~\cite{cai2011single} have shown that enlarging the DA can be achieved when RDTs are periodically well canceled. Further investigations~\cite{wei2023minimizing, wei2024analysis} revealed that even when perfect periodic cancellation is unattainable, minimizing their variations along the longitudinal direction $s$ is helpful. Therefore, it is of interest to compare the RDTs and their variations in the two configurations calculated with the code \texttt{MAD-X}~\cite{de2023jacow}. Although RDTs are completely ignored during chaos suppression, we observe that, under the CS approach, most third- and fourth-order RDTs -- as well as their variations -- are passively well minimized, as illustrated in Fig.~\ref{fig:h3h4} and Tab.~\ref{tab:h3h4}.
      
  \begin{table}[ht]
  \centering
  \begin{tabular*}{0.8\columnwidth}{@{\extracolsep{\fill}} c|r|r}
  \hline\hline
         & RDTM & CS \\
  \hline
        $h_{3000,0}$ & 8.71 & 9.72 \\
        $h_{2100,0}$ & 25.03 & 16.66  \\
        $h_{1020,0}$ & 14.24 & 13.17  \\
        $h_{1011,0}$ & 23.56 & 23.42  \\
        $h_{1002,0}$ & 11.36 & 10.77  \\
  \hline
        $h_{4000,0}$ &  117.41& 106.62 \\
        $h_{3100,0}$ &  324.10&  538.52\\
        $h_{2011,0}$ &  581.79&  779.13\\
        $h_{2002,0}$ &  653.31&  469.81\\
        $h_{2020,0}$ &  282.80&  416.77\\
        $h_{0031,0}$ &  448.32&  390.57\\
        $h_{0040,0}$ &  120.54&  117.33\\
        $h_{1120,0}$ &  725.11&  521.40\\
  \hline
  \end{tabular*}
  \caption{The averaged longitudinal variation of the third- and fourth-order RDTs}
  \label{tab:h3h4}
  \end{table}

\subsection{Chromaticity and local momentum aperture}

\begin{figure}
  \centering
  \includegraphics[width=0.9\columnwidth]{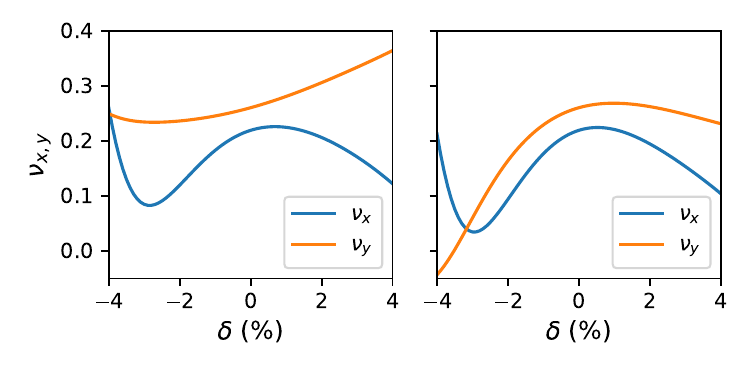}
  \caption{The chromaticity (energy-dependent detuning) for two configurations..} 
  \label{fig:chrom}
  \end{figure}

  Chromaticity describes the variation of the betatron tune with respect to the beam energy deviation ($\delta$). Although harmonic sextupoles do not affect the linear chromaticity, they influence higher-order terms that can drive off-momentum particles into resonance and ultimately limit the LMA. Controlling energy-dependent detuning is therefore also essential for stable operation. A comparison of the two configurations is shown in Fig.~\ref{fig:chrom}. Within the required range of $\delta \in [-2.5\%, 2.5\%]$, tune variation is well confined in both cases.
    
\begin{figure}
  \centering
  \includegraphics[width=0.9\columnwidth]{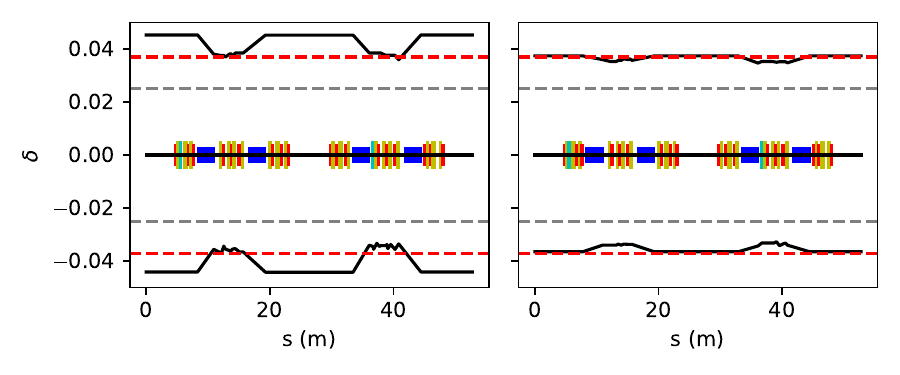}
  \caption{The LMAs (solid black line) within one supercell for two configurations. The dashed gray lines represent the desired momentum aperture at $\delta=\pm2.5\%$. The dashed red lines represent the RF acceptance provided by a $\SI{3.6}{MV}$ cavity voltage.} 
  \label{fig:lma}
  \end{figure}

  With a given chromaticity for the entire ring, the stability of off-momentum particles remains a localized property. At a given longitudinal position, the LMA defines the maximum relative energy deviation a particle can sustain while remaining stable~\cite{borland2000elegant}. In high-density bunched beams, the Touschek effect leads to intra-beam scattering, whereby particles exchange transverse momentum for longitudinal momentum. To ensure adequate beam lifetime, sufficient LMA is required for the majority of scattered particles to survive. Fig.~\ref{fig:lma} shows the comparison of LMA defined by the nonlinear optics within one cell.

  When the RF cavities are in operation, the Touschek lifetime is ultimately determined by the smaller of the RF acceptance and the LMA. At NSLS-II, the total RF voltage of $\SI{3.6}{MV}$ provides an RF acceptance comparable to that obtained with the SH4 sextupole family turned off. The lifetimes computed from the scattering loss distribution along the longitudinal direction for these two configurations are 13.2 and 10.6 hours, respectively. The reduction might be caused by the dip in the off-momentum horizontal DA at $\delta=2\%$ (though it remains larger than $\SI{15}{mm}$ in Fig.~\ref{fig:xd}), and the vertical tune's crossing of the integer line around $\delta=-3.5\%$ in Fig.~\ref{fig:chrom}. Nevertheless, this level of reduction is considered acceptable when the machine operates in top-off mode with an injection interval of 1 -- 2 minutes. Experimentally, both configurations were observed to perform comparably and satisfy the requirement of $\delta \in [-2.5\%,2.5\%]$, corresponding to a beam lifetime of approximately three hours.
    
\subsection{Sensitivity to errors}

\begin{figure}
  \centering
  \includegraphics[width=0.9\columnwidth]{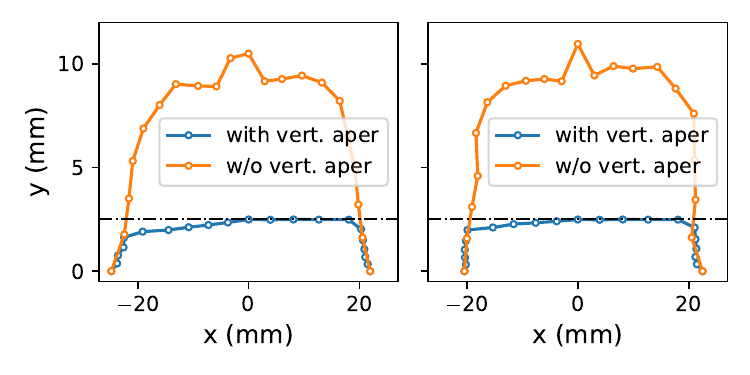}
  \caption{The DAs in the presence of higher-order multipole errors and physical apertures for two configurations. The dashed black lines denote in-vacuum undulators vertical gaps at $y=\pm \SI{2.5}{mm}$.} 
  \label{fig:da_err}
  \end{figure}

  The performance of the DA unavoidably degrades in the presence of magnetic field imperfections, such as undesired higher-order multipole components, and physical aperture constraints, particularly the tight vertical gaps of insertion devices (IDs) in light source rings. Such gaps -- typically on the order of a few millimeters -- can cause injected beam scraping due to nonlinear coupling. A robust lattice design is essential to ensure tolerance against realistic imperfections. Fig.~\ref{fig:da_err} illustrates that the shrunk DAs, resulting from specified multipole errors~\cite{dierker2007nsls} and restricted physical apertures~\cite{tanabe2015insertion}, are also comparable between the two configurations.

\subsection{Correlation with dynamic aperture}
  
  The ultimate goal of nonlinear lattice design is to achieve sufficient DA and LMA. However, DA simulations are computationally intensive and therefore impractical for large-scale ring design, while also providing limited physics-based insight into the underlying particle dynamics. A more effective approach is to employ computationally efficient methods -- such as RDTs or CIs -- to quickly filter out unsuitable candidates, followed by detailed tracking simulations applied selectively to the most promising configurations. Within this framework, understanding the correlation between the optimization objectives and the final DA is critical for achieving robust solutions. Reference~\cite{yang2011multiobjective} demonstrates that minimizing RDTs is a necessary but insufficient condition for obtaining a large DA; although some correlation exists, it is relatively weak. Interestingly, by applying a chaos suppression approach, we observed a stronger correlation, which led to high-quality solutions even when using fewer sextupole families.

\begin{figure}
  \centering
  \includegraphics[width=0.9\columnwidth]{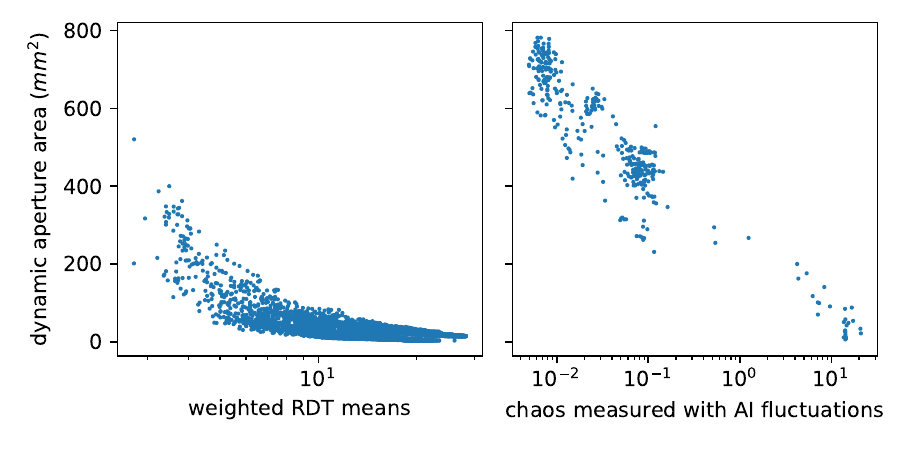}
  \caption{The correlation between the DA area and the two optimization metrics: the RDT metric (left subplot) and the CI (right subplot). Each data point represents a unique set of sextupole configurations generated during the optimization process.} 
  \label{fig:correlation}
  \end{figure}

  To further explore this, a numerical experiment was conducted to compare the correlation between DA and optimization objectives based on RDTs and CIs. The control variables were the same five families of sextupoles for both optimization strategies. For the RDT minimization, a multi-objective genetic algorithm optimizer was employed to simultaneously suppress multiple $h_{3,4}$ RDTs, including three ADD terms~\cite{wang2012explicit}. In such a case, a candidate is considered potentially good if it is not dominated by other solutions (i.e., if no other candidate has better values for all objectives). However, many of these candidates -- which may only have a very few small RDTs -- typically do not provide a good DA, yet they still survive the optimization process. When chaos suppression was used as a single objective for DA optimization, a simple downhill simplex algorithm was employed. Once good solutions are found, the algorithm continues to explore their neighboring region, generating many similarly good candidates. The correlation and the distribution of candidates reveal that using the CS approach is not only simpler (involving a single objective), but it also tends to yield a larger number of high-quality candidate solutions with an accelerated convergence, as shown in Fig.~\ref{fig:correlation}. In its left subplot, the RDT metric is an aggregate sum. To account for the dramatic variation in RDT magnitudes across different orders, each individual RDT was first weighted by the mean value of all candidates in the data pool before summation to establish the correlation~\cite{yang2011multiobjective}. A significantly stronger correlation is observed between the DA and the CI. Furthermore, this experiment explains why, in our early design stage, a six-family sextupole scheme was required when optimization targeted RDTs: with only five families, the optimal RDT solutions failed to provide the desired DA.

\subsection{Experimental validation}

\begin{figure}
  \centering
  \includegraphics[width=0.9\columnwidth]{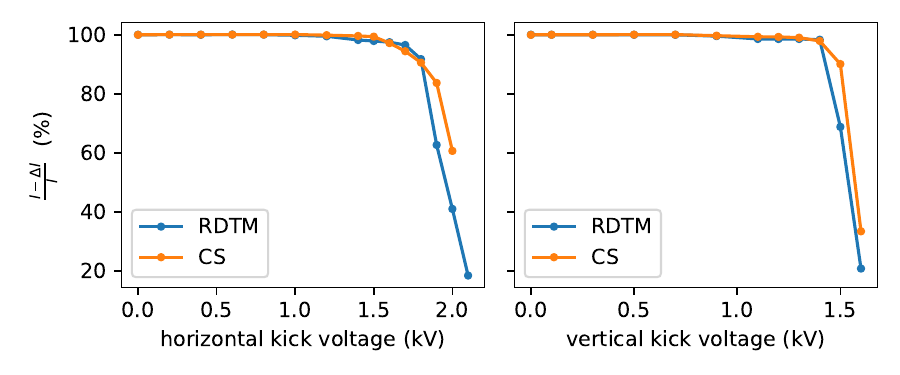}
  \caption{Comparison of measured DA in the horizontal (left) and vertical (right) planes. The beam loss pattern subjected to progressively increasing pinger kicks for the RDTM configuration is shown with the blue line-dots, while that for the CS configuration is shown in orange. The strong similarity in the loss patterns validates the comparable DA performance between the two configurations.} 
  \label{fig:measuredDA}
  \end{figure}

  The comparison was validated by measuring the DAs of these two sextupole configurations at the NSLS-II bare ring with all insertion device gaps open. Experimentally, DAs can be determined by observing stored beam loss following excitation with a pulsed magnet (commonly referred to as a pinger magnet). In this measurement, a short train of stored beam bunches was subjected to progressively increasing kick voltages, as illustrated in Fig.~\ref{fig:measuredDA}. The resulting beam loss patterns exhibit strong similarity, validating the comparable dynamic aperture performance observed in prior simulations. Additionally, the beam lifetime remains similar under identical bunch filling patterns and beam current conditions, which confirms that their LMAs are comparable as well.

\section{Summary}

  By suppressing chaotic behavior, we obtained and experimentally validated an alternative nonlinear lattice scheme for the NSLS-II storage ring. This CS approach enhances the beam's resistance to chaos and provides comparable DA and LMA performance while requiring fewer sextupoles than the conventional RDTM design. This finding challenges the traditional approach, which places excessive emphasis on minimizing ADD, an objective that our results suggest is less critical than previously assumed. Furthermore, we demonstrated a significantly stronger correlation between the DA area and the CI metric compared to the RDT metric, confirming that targeting global chaos is a more robust and efficient strategy for nonlinear lattice optimization.
  
  Another method for DA optimization, though not the primary focus of this paper, is directly based on long-term tracking simulations~\cite{borland2009direct}. Although this approach could also yield high-quality solutions when sufficient computational resources are available, it often provides limited physical insight into the underlying particle dynamics. While such tracking-based studies have been conducted at NSLS-II~\cite{yang2011multiobjective}, the emphasis of this paper is on physics-guided approaches that elucidate the underlying mechanisms of nonlinear dynamics, thereby ensuring robust and generalized design principles.    

\begin{acknowledgments}

  We thank D.~Xu (BNL), Y.~Hao (MSU), S.~Nagaitsev (BNL), X.~Huang (SLAC), and G.~Hoffstaetter (Cornell University) for their valuable discussions. This research is supported by the U.S. Department of Energy under Contract No.~DE-SC0012704 and Field Work Proposal 2025-BNL-PS040.

\end{acknowledgments}

\bibliography{ref.bib}

\end{document}